\documentclass[journal]{IEEEtran}
\usepackage{amsmath,amsfonts}
\usepackage{algorithmic}
\usepackage{algorithm}
\usepackage{array}
\usepackage[caption=false,font=normalsize,labelfont=sf,textfont=sf]{subfig}
\usepackage{textcomp}
\usepackage{stfloats}
\usepackage{url}
\usepackage{verbatim}
\usepackage{graphicx}
\usepackage{cite}
\usepackage{multirow}
\usepackage{threeparttable}
\usepackage[most]{tcolorbox}

\begin{document}

\title{Target Speech Diarization with Multimodal Prompts}

\author{Yidi Jiang,~\IEEEmembership{Student Member, IEEE}, Ruijie Tao,~\IEEEmembership{Member, IEEE}, Zhengyang Chen,~\IEEEmembership{Student Member, IEEE}, \\ Yanmin Qian,~\IEEEmembership{Senior Member, IEEE} and Haizhou Li,~\IEEEmembership{Fellow, IEEE}

\thanks{Yidi Jiang and Ruijie Tao are with the Department of Electrical and Computer Engineering, National University of Singapore, Singapore 117583 (e-mail:yidi\_jiang@u.nus.edu; ruijie@nus.edu.sg).~\textit{(Corresponding author: Ruijie Tao.)}

Zhengyang Chen and Yanmin Qian are with the Shanghai Jiao Tong University, Shanghai 200240, China (e-mail: zhengyang.chen@sjtu.edu.cn; yanminqian@sjtu.edu.cn). 

Haizhou Li is with the Shenzhen Research Institute of Big Data, School of Data Science, The Chinese University of Hong Kong, Shenzhen 518172, China, and also with the Kriston AI, Xiamen 361026, China (e-mail: haizhouli@cuhk.edu.cn).
}
}

\markboth{Journal of \LaTeX\ Class Files,~Vol.~14, No.~8, August~2021}%
{Shell \MakeLowercase{\textit{et al.}}: A Sample Article Using IEEEtran.cls for IEEE Journals}


\maketitle

\begin{abstract}
Traditional speaker diarization seeks to detect  ``who spoke when'' according to speaker characteristics. Extending to target speech diarization, we detect ``when target event occurs'' according to the semantic characteristics of speech.  We propose a novel Multimodal Target Speech Diarization (MM-TSD) framework, which accommodates diverse and multi-modal prompts to specify target events in a flexible and user-friendly manner, including semantic language description, pre-enrolled speech, pre-registered face image, and audio-language logical prompts. 
We further propose a voice-face aligner module to project human voice and face representation into a shared space.
We develop a multi-modal dataset based on VoxCeleb2 for MM-TSD training and evaluation. Additionally, we conduct comparative analysis and ablation studies for each category of prompts to validate the efficacy of each component in the proposed framework. Furthermore, our framework demonstrates versatility in performing various signal processing tasks, including speaker diarization and overlap speech detection, using task-specific prompts. MM-TSD achieves robust and comparable performance as a unified system compared to specialized models. Moreover, MM-TSD shows capability to handle complex conversations for real-world dataset.
\end{abstract}


\vspace{-1em}
\section{Introduction}
\IEEEPARstart{H}{umans} have the ability to selectively attend to a specific sound source in a complex acoustic environment, that is commonly referred to as the cocktail party effect~\cite{cherry1953some}. Benefit from this remarkable auditory attention mechanism, human can effectively focus on a particular speaker of interest ~\cite{fritz2007auditory, mesgarani2012selective} since each speaker has the unique voice characteristic. The speaker diarization task aims to segment multi-talker speech based on speaker identities and determines ``who spoke when''~\cite{cheng2023target,medennikov2020target,czy-taslp}, which serves as a front-end for various downstream speech-related tasks.

In addition to speaker identity~\cite{desplanques2020ecapa,10497864}, there is interest in other semantic aspects of human speech (referred as ``target events'')~\cite{wang2023speech,tzinis2022heterogeneous,lebourdais2022overlapped}, such as male/female speech, multi-talker speech mixture, or the speech of a keynote speaker who speaks the most in a meeting. This indicates the need for determining target speech in a comprehensive and multi-dimensional manner. Therefore, we proposed a new paradigm termed ``target speech diarization'', which aims to identify ``when target event occurred'' guided by prompts specifying target events. Our previous work~\cite{jiang2023prompt} provided a feasible method that leverages prompt vectors to offer conditional information for specific target events. By switching between different prompt vectors, the framework can identify the corresponding event regions within an audio signal.

\begin{figure}[!t]
    \centering
      \includegraphics[scale=0.55]{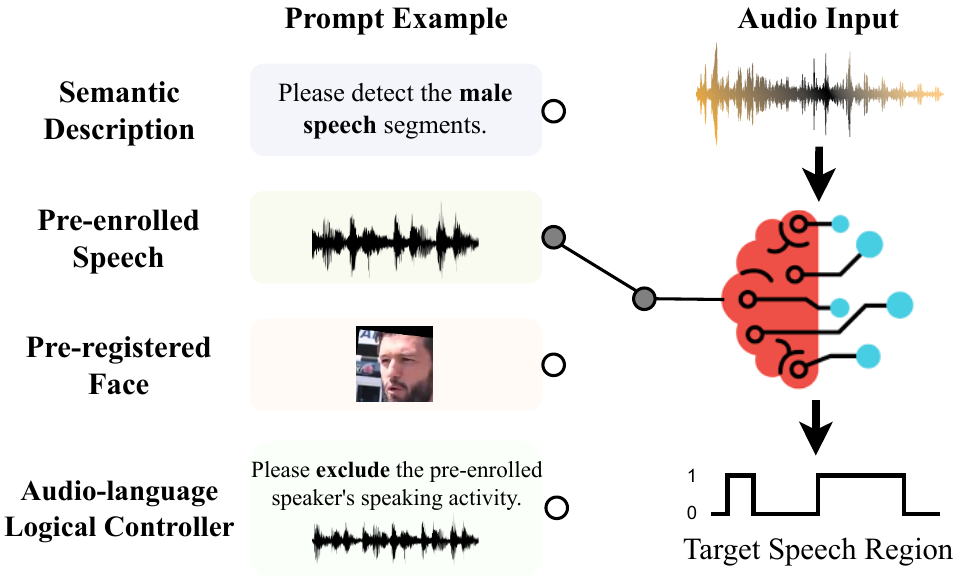}
      \vspace{-1.5em}
        \caption{The illustration of four types of prompts supported by our Multimodal Target Speech Diarization (MM-TSD) framework. The unified target speech diarization model can accommodate multi-modal and diverse prompts, including semantic language description, pre-enrolled speech and pre-registered face of the target speaker and the audio-text logical controller. Our framework then detects the activity regions of the target speech specified by the prompt. }
    \label{fig:illustration}
    \vspace{-1.2em}
\end{figure}

In real-world scenarios, the utilization of prompt vectors is constrained and not user-friendly enough. Humans perceive and interact with the world through multiple modalities, including linguistic, auditory, and visual cues. For instance, in a meeting scenario, we rely on language instructions, spoken speech, facial expressions and gestures to communicate. Similarly, prompts specifying target events can exist in various multi-modal formats such as language instructions, pre-enrolled speech and pre-registered faces, or even the combination of them.
However, it remains a challenge how to integrate multi-modal prompts for diverse scenarios into a system.
We hypothesize that it's feasible to project multi-modal prompts describing the semantic characteristics of speech into a shared semantic space. This projection would facilitate the processing of multi-modal information within a unified model and interaction between speech input and multi-modal prompts.
Therefore, our objective is to develop a unified framework that allows flexible user interaction through a range of multi-modal prompts, allowing users to identify target events across multi-dimensional semantic aspects according to their requirements and preferences.

In ``target speech diarization'' task, one uses pre-enrolled speech as a prompt to specify a desired target event, just like the pre-enrolled speech for a target speaker in target speech voice activity detection (TS-VAD)~\cite{medennikov2020target,cheng2023target}, where speech regions that correspond to a reference speaker are detected. 

However, it is challenging to describe complex concepts with pre-enrolled speech, such as overlapping speech or the most talkative speaker. Moreover, pre-enrolled speech is not always available. Natural language, as the most natural way of human communication, is commonly used to express complex concepts. Therefore, it serves as a natural choice of human prompt. Prior works have focused on language-queried audio source separation~\cite{liu2022separate,liu2023separate} and text-guided target speaker extraction~\cite{hao2023typing}, which utilize the natural language to achieve conditional separation and extraction functions. 
Nevertheless, building a target speech diarization system that can effectively handle the complex and various natural language expressions poses challenges in modeling intricate audio-text interactions. For instance, the same speech event can be described through various text prompts, such as ``male speech'' and ``the voice of the man''.
The system must be able to correlate these diverse text prompts with the same speech event and subsequently identify the corresponding regions within the audio input.

Apart from using natural language to specify a particular speech event, recent studies~\cite{tao2021someone,jiang23c_interspeech,cheng2024multi} have demonstrated the effectiveness of incorporating lip movements to detect speech of a particular speaker through audio-visual synchronization. While these methods are always limited to scenarios with the high frame rate video. Furthermore, many works~\cite{chung2020perfect,nagrani2018seeing,nagrani2018learnable,oh2019speech2face} have also explored the usage of still face images and verified the association between voice representations and face appearance due to shared latent factors, such as age, gender and nationality/accent. In the field of audio-visual speech processing, FaceFilter~\cite{chung2020facefilter} has explored audio-visual speech separation conditioned on a still face image of a target speaker. In this work, we make the first attempt for face-based target speech diarization using static face image as prompt to specify the desired speaker.

As previously discussed, while single-modal prompts can identify specific speech event in audio signals, real-world communication often involves combining multiple modalities to convey concepts.
We also explore the interactive multi-modal prompts for complex logical operations. For example, in certain scenarios, there is a need to filter out the target speaker's voice. By inputting the command ``Exclude the pre-enrolled speaker's speech" along with a reference speech, we can identify segments excluding the enrolled speaker. In this study, we explore the interaction between audio and text prompts, where text commands serve as a logical controller to determine whether detect or exclude the pre-enrolled speaker's voice. With the integration of ``exclusion" related commands, our framework functions as a ``NOT Gate'', allowing precise exclusion of the target speaker's voice when necessary.

As shown in Figure~\ref{fig:illustration}, in this work, our primary objective is to establish a unified target speech diarization model capable of accommodating multi-modal prompts to specify various target events. These prompts include semantic language description, pre-enrolled speech, pre-registered face image, and audio-language logical prompts. 
We propose Multimodal Target Speech Diarization~(MM-TSD) framework, which contains the modality-specific prompt encoders and modality-agnostic Transformer encoder-decoder for handling the diverse application scenarios.

This work is an extension of our previous study, which was presented at ICASSP~\cite{jiang2023prompt}. Our contributions in this work are as follows:
\begin{itemize}
    \item {To the best of our knowledge, the proposed MM-TSD is the first attempt for end-to-end target speech diarization, supporting audio, visual, textual and audio-text multi-modal prompts to specify the target speech events. This work sets a reference benchmark and provides valuable insights into multi-modal prompt-guided target speech processing.}
    \item {We introduced the use of static facial cues in diarization-related tasks and proposed a voice-face aligner module to establish correspondence between human face and voice biometrics.}
    \item {We evaluated MM-TSD framework across various modalities of prompts and semantic attributes on both simulated and real-world datasets to show its effectiveness in detecting the prompt-specified target events. The evaluations further confirm our hypothesis that MM-TSD can project the multi-modal prompts and speech input into a shared space within a unified framework.}
\end{itemize}

\vspace{-0.8em}
\section{Related Work}
\subsection{Speaker Diarization}
Speaker diarization seeks to delimit the boundaries of speaker turns in a multi-talker speech according to speaker characteristics, i.e. voiceprint. Taking advantage of speaker-specific information~\cite{jiang23c_interspeech}, target speaker voice activity detection (TS-VAD)~\cite{medennikov2020target} employs speaker embeddings of speakers during the diarization process.

To obtain the voiceprint of all speakers in the conversation, TS-VAD system employs an additional clustering-based diarization system to identify the single-speaker segments for each individual. Subsequently, a pre-trained speaker embedding extractor is utilized to obtain speaker embeddings. TS-VAD applies these speaker embeddings as the references to guide the diarization process and detect the speaking status of each speaker~\cite{cheng2023target,wang2023target}. Inspired by the success of TS-VAD, our framework utilizes the pre-enrolled speech utterance as the prompt to detect the speaking activities of the target event.

\vspace{-0.5em}
\subsection{Language-Queried Audio Processing}
In recent years, audio-language processing has emerged as a novel research domain. Introducing text modality into speech tasks provides precise descriptions and guidance, offering significant value across various application scenarios in a user-friendly manner. For instance,~\cite{liu2022separate,liu2023separate} propose language-guided audio source separation, which aims to isolate specific sources from audio mixtures using natural language queries. 
Furthermore, language-based audio retrieval~\cite{mei2022language,koepke2022audio,lou2022audio} has proven efficacy for multimedia content retrieval and sound analysis. 
In the context of our proposed target speech diarization, the text modality can clearly define the target speech event for flexible user interaction. Motivated by that, we would like to construct a TSD system that can respond to diverse language-based queries.

\vspace{-0.8em}
\subsection{Voice-Face Biometric Matching}
In addition to text and audio, humans also rely on visual cues to perceive and interact with the world, which motivates the incorporation of visual cues into the TSD process.
A straightforward approach is to employ lip movements of a particular speaker as visual cues~\cite{tao2021someone,pan2021muse}. While such audio-visual models have demonstrated remarkable outcomes, they often require high-quality video data and high computational resource, which may not always be available in real-world application.

Recognizing this limitation, employing a single static face image as a visual cue for speech processing is an alternative~\cite{chung2020facefilter}. Each speaker has distinct voice characteristics, including the vocal tract shape, pitch, and prosody variation, as well as unique facial landmarks. Recent research demonstrated that voice representations can exhibit correlations with face appearance due to shared latent factors such as age, gender, and ethnicity/accent~\cite{thornhill1997developmental,hollien1960measurements}. This connection can be leveraged in target speech diarization task to specify the target event, i.e., the speech corresponding to a given static face image. While this connection may not be robust among speakers of the same gender, nationality or age range, the exploration for face-based speech processing still remains meaningful. 
From a practical standpoint, users' profile images are always accessible on various mobile devices, social networks, and company groupware, enhancing the accessibility of such audio-visual solutions.
Our research represents the first investigation to utilize the static face image as the visual prompt in target speech diarization, leveraging the inherent voice-face correlation.

\section{Task Formulation: Target Speech Diarization}
\label{ssec:task_formulation}
In this section, we outline the formulation of our proposed target speech diarization task.
First, we introduced two key concepts for our task: semantic attribute and semantic value~\cite{jiang2023prompt}. Semantic attributes contain a set of speech properties such as speaker identity and gender, which represent the criteria of demarcating speech segments. Each semantic attribute takes on one or multiple semantic values associated with specific events. For examples, in speaker diarization task, speaker identity is the semantic attribute. The specific speaker ID is semantic value and his/her speaking region is its aligned speech event. 

In the target speech diarization system, it simultaneously takes audio and the prompt that specify the target speech event as inputs and outputs the corresponding target event regions. For example, when provided with pre-enrolled speech~(prompt) of ``Speaker A''~(semantic value), the framework will output the speaking regions of ``Speaker A''~(target event).

The core of a TSD system lies in the prompt, which specifies the target event and guides the TSD process. As depicted in Figure~\ref{fig:illustration}, prompts are available in various formats for different application scenarios.

\begin{figure*}[!t]
    \centering
      \includegraphics[scale=0.9]{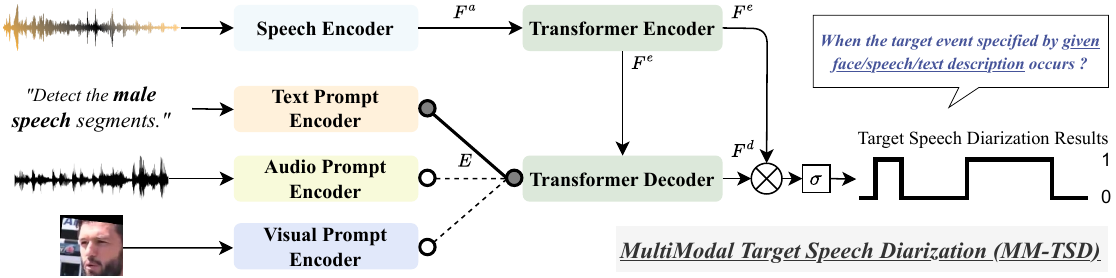}
    \caption{Our MM-TSD framework takes an audio signal and a switchable multi-modal prompt as inputs, to output frame-wise binary classification of the prompt-specified speech event. 
    It accommodates diverse prompt types such as semantic language descriptions, pre-enrolled speech, pre-registered face images or the combination of audio-text logical prompts to specify the target event. This framework comprises a speech encoder, three modality-specific prompt encoders and a Transformer encoder-decoder structure. Then a dot product $\bigotimes$ is applied between the encoder and decoder outputs, followed by a sigmoid operation $\sigma$ to calculate the target event occurrence probability at each frame.} 
    \label{fig:overview}
    \vspace{-0.7em}
\end{figure*}

\subsubsection{TSD with Semantic Description}
\label{subsubsection:textual tsd}
Humans perceive audio signals based on the distinguishing semantic characteristics, such as female speech or non-overlap speech in the audio signal. 
This scenario enables users to incorporate such perceptual cues as text-based semantic descriptions to guide target speech diarization. For example, when semantic attribute is gender, there are two semantic values: female and male. The prompt format is natural language, for example, ``please detect the female speech regions". Then the target event is the female speech segments.

\subsubsection{TSD with Pre-enrolled Speech}
In this scenario, the semantic attribute is speaker identity, the semantic value is a specific speaker ID, the prompt format is the pre-enrolled speech of the target speaker, and the target event is the speaking regions of the target speaker. 
The system aims to detect the speaking activity of each prompt-specified speaker.

\subsubsection{TSD with Pre-registered Face}
Similarly, in this scenario, the semantic attribute is face identity, and the semantic value is still a specific speaker ID. The prompt format is the pre-registered face to specify the target speaker, and the target event is the speaking regions of the specified speaker. 
This scenario offers users the capability to identify speaking regions of interest by providing the system with the pre-registered face of the target speaker.

\subsubsection{TSD with Audio-Language Logical Controller}
\label{subsubsection:audio-text}
In this scenario, there are two related attributes: ``included identity'' and ``excluded identity''.
The prompts consist of natural language serving as a logical controller and pre-enrolled speech to specify the target speaker. For instance, when the semantic attribute and value are ``excluded identity'' and ``Speaker A'', respectively, the target event is the regions where the voice of ``Speaker A'' doesn't occur. 
This scenario offers users the flexibility to decide whether to detect or exclude the pre-enrolled speaker's active regions within the audio mixture. 

\vspace{-0.5em}
\section{Multimodal Target Speech Diarization}
As illustrated in Figure~\ref{fig:overview}, the proposed MM-TSD system consists of a speech encoder, multi-modal prompt encoders and a Transformer encoder-decoder structure. Our framework is designed to flexibly switch and accommodate one or multiple prompts simultaneously, outputting the associated target event(s) regions accordingly.

Firstly, the speech encoder is employed to extract the audio feature sequence from the speech input, denoted as $F^a$. 
Concurrently, a prompt which lies in various modalities is used to specify the target event. 
To process input from different modalities, we employ three modality-specific prompt encoders: a text prompt encoder, an audio prompt encoder and a visual prompt encoder. These encoders convert multi-modal prompts~(text command, pre-enrolled speech or pre-registered face) into corresponding prompt embeddings
$E^T$, $E^A$ and $E^V$, respectively. Each prompt embedding is characterized by a dimension $D$. To ensure reliable modality representations, we apply pre-training techniques for each modality encoder.

Then, to align the prompt-specified event with the input speech, the Transformer encoder-decoder takes $F^a$ and prompt embedding $E$ as inputs and outputs the prediction sequence $\hat{\mathrm{Y}} \in (0, 1)^{1 \times T}$, where $T$ represents the number of frames. The values of $\hat{\mathrm{Y}}$ denote the target event occurrence probability at each frame. Specifically, the Transformer encoder receives $F^a$ and outputs the frame-level speech representation $F^e = [F^e_1, F^e_2, \ldots, F^e_T] \in \mathbb{R}^{T \times D}$. The Transformer decoder takes prompt embeddings $E$ and $F^e$ as inputs, and outputs $F^d$ with dimension $D$. Finally, we performed a dot product operation between the decoder output $F^d$ and the encoder output $F^e$ and applied a sigmoid operation to get the prediction sequence $\hat{\mathrm{Y}}$. 

\vspace{-0.8em}
\subsection{Speech Encoder}
To obtain the robust representation of the input speech, in our framework, we employ a pre-trained WavLM encoder~\cite{chen2022wavlm} as the speech encoder to obtain speech representations $F^a$. The WavLM encoder was designed to learn universal speech representations from vast amounts of unlabeled speech data, ensuring the universality and robustness of the frame-level audio representations. With consideration for the trade-off between computational efficiency and speech information, we utilized its convolutional feature encoder and the first three layers of the Transformer encoder, freezing them during our training process.


\vspace{-0.8em}
\subsection{Text Prompt Encoder with LoRA}
The goal of the text prompt encoder is to ensure that our framework can accommodate diverse textual descriptions for each target speech event. Specially, the text encoder is designed to map various sentence descriptions, which specify the same event, into a similar embedding space. To achieve this, we utilize a Pre-trained Language Model~(PLM) as the text prompt encoder to extract prompt embedding. Additionally, we explore a lightweight fine-tuning approach to achieve the training efficiency and adaptation.

\begin{figure}[!t]
    \centering
    \includegraphics[scale=0.7]{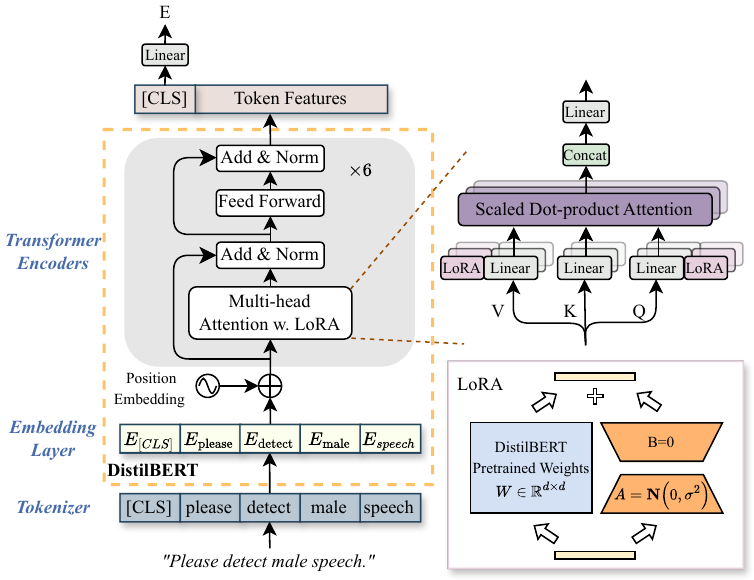}
    \caption{Text prompt encoding. The textual prompt is first processed by a tokenizer to generate word tokens, including a ``[CLS]'' token at the beginning. We utilize a pre-trained DistilBERT encoder with Low-Rank Adaptation (LoRA) to derive the sentence embedding. The feature of the ``[CLS]'' token is then used as the prompt embedding $E$.}
    \label{fig:text_encode}
    \vspace{-1.em}
\end{figure}

As decipted in Figure~\ref{fig:text_encode}, the text prompt is firstly tokenized using the BERT~\cite{devlin2018bert} Tokenizer, converting the textual command to tokens.
The tokens are then fed into the PLM text encoder.
To optimize training efficiency, we employ the DistilBERT model~\cite{sanh2020distilbert} as our PLM text encoder. DistilBERT is a fast, cost-effective, and lightweight Transformer model derived from the distillation process of the BERT base model, which use offers fewer parameters but preserve over 95\% of BERT's performance~\footnote{\url{https://huggingface.co/docs/transformers/en/model_doc/distilbert}}. 

The DistilBERT model begins with an embedding layer that transforms tokens into the token embeddings, incorporating position embeddings, and proceeds through six Transformer encoder layers to generate the sentence's hidden state embedding. After that, we regard the ``[CLS]'' token embedding as a pooled embedding with condensed semantic information, and pass it through a linear layer followed by ReLU and dropout.
Then, we use another linear layer to adapt the embedding to the dimension $D$, serving as our text prompt embedding $E^T$.

Furthermore, to adapt the PLM to our textual prompt semantic space without conducting full fine-tuning on the PLM text encoder, we adopt the parameter-efficient Low-Rank Adaptation (LoRA) technique~\cite{hu2022lora}. This approach involves incorporating trainable rank decomposition matrices into each layer of the Transformer architecture. Consequently, PLM adaptation can be achieved with a reduced number of trainable parameters. Specially, the LoRA structure is incorporated into query and value linear layers in each multi-head attention of every Transformer layer.
\vspace{-0.8em}
\subsection{Audio Prompt Encoder}
\label{subsection:audio prompt encoder}
The audio prompt is the pre-enrolled speech of the target speaker. To guarantee robust performance, we leverage a pre-trained ECAPA-TDNN~\cite{desplanques2020ecapa} speaker recognition model to obtain the target speaker embedding as audio prompt embedding $E^A$. The ECAPA-TDNN model has demonstrated reliable performance in speaker recognition tasks.

The ECAPA-TDNN model employs emphasized channel attention to selectively focus on critical parts of the speech signal, propagating that information through the network and aggregating it to make a final decision. From the variable lengths of input utterances, the output speaker embedding has the fixed dimension $D$. We freeze the parameters of this module in our framework, since our purpose is to obtain robust embedding for the target speaker.

\vspace{-0.8em}
\subsection{Visual Prompt Encoder with Voice-Face Aligner}
Previous works have demonstrated a correlation between the facial appearance and voice characteristics~\cite{chung2020perfect,nagrani2018seeing,nagrani2018learnable,oh2019speech2face}. Building on this cross-modal association, we investigate using pre-registered face images of the target individual as prompts to identify their speech regions. In MM-TSD, we employ a pre-trained ResNet50 model~\cite{he2016deep} as the face prompt encoder, known for its robust face recognition performance trained on large-scale face datasets. However, the face embeddings extracted from the pre-trained model may exist in a mismatched space with input audio representations. 

\begin{figure}[!ht]
    \centering
      \includegraphics[scale=0.85]{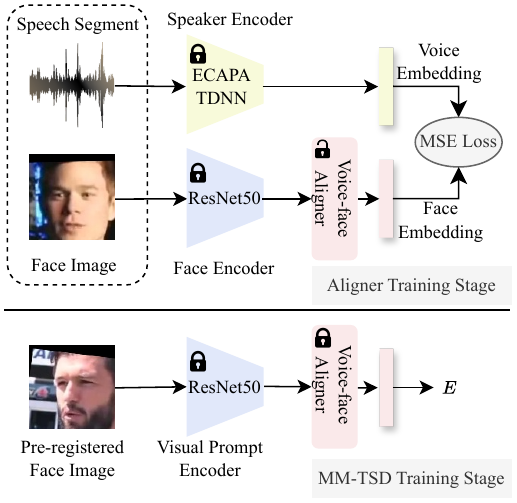}
      \vspace{-0.5em}
    \caption{Voice-face alignment involves inputs from speech segments and face images belonging to the same individual, which are denoted within dashed boxes. 
    We utilize pre-trained ECAPA-TDNN as the speaker encoder and ResNet50 as the face encoder to extract respective embeddings from the speech segment and face image. Following this, a voice-face aligner is employed to match face identity with voice characteristics in a shared embedding space. During the aligner training phase, the voice-face aligner is trained using Mean Squared Error (MSE) loss. In the subsequent MM-TSD training phase, the visual prompt encoder and voice-face aligner are both frozen to derive the visual prompt embedding $E$. }
    \label{fig:align}
    \vspace{-0.5em}
\end{figure}

To address this discrepancy and leverage the intrinsic associations between human face and voice biometrics, we incorporate a novel voice-face aligner module with the additional ``Aligner Training Stage'', as shown in Figure~\ref{fig:align}. 
The voice-face aligner is designed to learn associations between audio and visual inputs, encompassing general identity features~(such as gender, age, and ethnicity) and appearance features~(such as prominent facial attributes like big nose, chubby cheeks, or double chin). Its goal is to establish correspondence between voice-face identity pairs and bridge the modality gap between voice and face embeddings.

The aligner training pipeline is illustrated in Figure~\ref{fig:align}. 
Each training data sample consists of the facial image and the corresponding reference speech from the same individual, which offer biometric information from diverse perspectives. 
Voice embeddings are obtained using a pre-trained speaker encoder, ECAPA-TDNN, similar to the audio prompt encoder introduced in Section~\ref{subsection:audio prompt encoder}. The face embedding is extracted from a pre-trained face encoder and a trainable voice-face aligner. This voice-face aligner is trained using Mean Squared Error~(MSE) loss, which quantifies the probability that the voice and face embeddings belong to the same person.

Finally, after the voice-face ``Aligner Training Stage'', we fix the ResNet50 face encoder and voice-face aligner for the entire ``MM-TSD Training Stage''. These two modules cooperate with each other to generate aligned face embeddings, serving as our visual prompt embeddings $E^V$.

\begin{figure}[!t]
    \centering
      \includegraphics[scale=0.8]{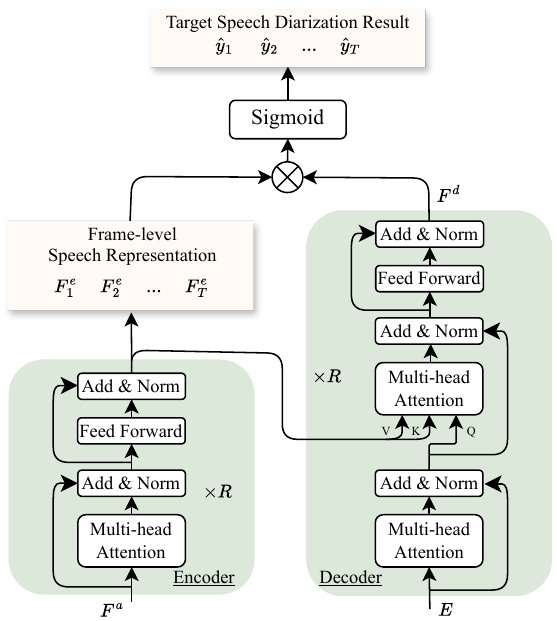}
    \vspace{-0.5em}
    \caption{The encoder receives the speech embedding $F^a$, which is extracted from the speech encoder, and produces a frame-level speech representation $F^e$. The decoder utilizes the prompt embedding $E$ as the query within a cross-attention mechanism, with $F^e$ serving as both the key and value. This setup enables precise alignment and interaction between the speech embedding $F^a$ and prompt embedding $E$ to detect the prompt-specified target event activities within the speech signal. $\bigotimes$ denotes the dot product operation between transformer encoder and decoder ouputs.}
    \label{fig:transformer}
    \vspace{-1em}
\end{figure}

\vspace{-0.8em}
\subsection{Transformer Encoder-Decoder} 
After obtaining the speech representation $F^a$ from the speech encoder and the prompt embedding $E$ from the prompt encoders, we feed them into a Transformer~\cite{vaswani2017attention} encoder-decoder architecture to predict the MM-TSD output sequence. This design leverages the self-attention and cross-attention mechanisms to capture intricate temporal patterns in the audio data and align relevant information with the prompt embeddings which serve as query. 

Within the Transformer encoder, self-attention enables interaction among the learnt speech representations, enhancing the overall quality of frame-level speech representations. As shown in Figure~\ref{fig:transformer}, the Transformer encoder produces encoder memory $F^e$ as frame-level speech representation. Both the prompt embedding $E \in \mathbb{R}^{1 \times D}$ and the Transformer encoder memory $F^e \in \mathbb{R}^{T \times D}$ are then fed into the Transformer decoder. Here, the prompt embedding $E$ serves as the query in the cross-attention structure, while the encoder memory $F^e$ serves as the key and value.
The cross-attention module within the Transformer decoder enables prompt embeddings to attend to all frame-level speech representations, ensuring that the resulting decoder outputs capture the most relevant information about the prompt-specified target event.


With the Transformer encoder memory and decoder output, we can calculate the posterior probability that each frame belongs to the prompt-specified event through a simple dot product operation:

\vspace{-0.5em}
\begin{equation}
    \hat{Y} = \sigma(F^d {F^e}^\top) \in (0,1)^{1 \times T}
\end{equation}
where the $\sigma$ symbol corresponds to the element-wise sigmoid function. MM-TSD benefits from this Transformer encoder-decoder to accurately identify and detect target event regions based on the prompts, achieving a robust and adaptable solution for our task.

\vspace{-1.em}
\subsection{Loss Function}
The learning targets of our framework are frame-wise binary ground truth labels $Y \in \{0,1\}^{1 \times T} $ of the target event.
We utilized binary cross-entropy loss to train our model, as defined in Equation~\ref{eq:loss}. $\hat{y}_t$ and $y_t$ represent the predicted and ground-truth labels of the specific target event for the $t^{th}$ audio frame, where $t \in [1, T]$. The loss function is designed to minimize the difference between predicted and ground-truth labels, encouraging our model to accurately detect target event activities.
\begin{equation}
    \mathcal{L} = -\frac{1}{T} \sum_{t=1}^{T} [ (y_t \cdot log \hat{y}_t) + (1-y_t) \cdot log (1- \hat{y}_t) ]
    \label{eq:loss}
\end{equation}

\vspace{-0.5em}
\section{MM-TSD Benchmark}
In this section, we establish a comprehensive benchmark for the proposed target speech diarization task. It includes the dataset corpus design, text prompt formulation, and evaluation metrics. This benchmark serves as the basis for our experiments and provides a standard reference for future studies to ensure fair comparisons.

\vspace{-1.em}
\subsection{Data Corpus}
\subsubsection{Simulation Dataset for MM-TSD Training}
Since real-world speech datasets cannot meet all the required ground-truth labels according to application scenarios introduced in Section \ref{ssec:task_formulation}, we followed the recipe\footnote{https://github.com/BUTSpeechFIT/EEND\_dataprep/} proposed in~\cite{landini2023multi} to generate simulation datasets for MM-TSD training from VoxCeleb2~\cite{Chung18b} which is an audio-visual dataset derived from YouTube interviews. 
In~\cite{landini2023multi}, the authors proposed a method that leverages the statistics from real recordings to guide the synthesis of simulated data.
To create datasets that closely resemble real-world conversations, we utilized conversation statistics from the DIHARD II development set~\cite{ryant2019second} to generate 1000 hours of audio for each MM-2spk, MM-3spk, and MM-4spk dataset to simulate the condition with different number of speakers as details shown in Table~\ref{tab:sim data}. 

To showcase the generalization capabilities of our system, we have devised both ``Seen-Heard'' and ``Unseen-Unheard'' test sets. The ``Seen-Heard'' set comprises speakers present in the training data, while the ``Unseen-unheard'' set consists of entirely new speakers.
After filtering the heavily noisy and unavailable videos, we select 5,702 speakers in VoxCeleb2 for training purposes. Additionally, we reserve 49 speakers for unseen validation and another 65 speakers for the unseen test set. This setup allows us to assess our system's performance in scenarios where it encounters entirely new speakers.

\subsubsection{Simulation Dataset for MM-TSD with Audio Prompt Analysis}
To demonstrate the effectiveness of MM-TSD with audio prompts, we conduct comparison experiments with state-of-the-art speaker diarization systems. It follows the traditional two-stage speaker diarization training process: pre-training stage with the simulated dataset and adaptation stage with the real-world dataset.
To ensure a fair comparison, we followed the simulation configuration and method outlined in~\cite{czy-taslp} to generate two subsets: Audio-2spk and Audio-3spk, with the statistics from Part1 of the CALLHOME dataset.
As shown in Table~\ref{tab:sim data}, Audio-2spk and Audio-3spk has 2481 hours and 4226 hours training data, respectively, featuring utterances with 2 and 3 speakers each. The utterances in each subset have a fixed number of speakers.

\begin{table}[!ht]
\vspace{-1.em}
\caption{Simulated dataset configuration. \# Spk and \# Utt represent the number of speakers and utterances, respectively. Ovl.(\%) corresponds to the overlap ratio.}
\label{tab:sim data}
\centering
\begin{tabular}{m{1.3cm}<{\centering}m{1.6cm}<{\centering}m{0.8cm}<{\centering}m{0.7cm}<{\centering}m{0.7cm}<{\centering}m{1.3cm}<{\centering}}
\hline
Dataset & Real-world Data Statistic & Split & \# Spk & \# Utt & Duration~(hrs) \\
\hline
\multirow{2}{*}{MM-2spk} & \multirow{2}{*}{DIHARD II dev} & Train & 2 & 14,361 & 1,000 \\
&& Test & 2 & 145 & 11 \\
\hline
\multirow{2}{*}{MM-3spk} & \multirow{2}{*}{DIHARD II dev} & Train & 3 & 9,752 & 1,000 \\
&& Test & 3 & 96 & 11 \\
\hline
\multirow{2}{*}{MM-4spk} & \multirow{2}{*}{DIHARD II dev} & Train & 4 & 7,472  & 1,000 \\
&& Test & 4 & 71 & 10 \\
\hline
\hline


\multirow{2}{*}{Audio-2spk} & \multirow{2}{*}{\begin{tabular}[c]{@{}l@{}}CALLHOME \\ (Part1 2spk)\end{tabular}} &Train & 2 & 24,343 & 2,481 \\
&&Test & 2 & 118 & 12 \\
\hline
\multirow{2}{*}{Audio-3spk} & \multirow{2}{*}{\begin{tabular}[c]{@{}l@{}}CALLHOME \\ (Part1 3spk)\end{tabular}} &Train & 3 & 29,297 & 4,226  \\
&&Test & 3 & 86 & 12 \\
\hline
\end{tabular}
\end{table}

\subsubsection{Real-world Dataset}
The datasets with real recordings used in our experiments are presented in Table~\ref{tab:real data}.
We employed the CALLHOME~\cite{callhome} dataset for analyzing MM-TSD with audio prompts and the DIHARD II~\cite{ryant2019second} dataset for analyzing MM-TSD with text prompts. 

The CALLHOME dataset is divided into two parts according to the kaldi recipe~\footnote{https://github.com/kaldi-asr/kaldi/tree/master/egs/callhome\_diarization/v2}. Part 1 is used for model adaptation, while Part 2 is used for evaluation. 
We selected the best-performing models trained on the Audio-2spk and Audio-3spk Train sets, which were evaluated on the corresponding Test sets. Then we employed the model for finetuning adaption on CALLHOME Part 1 subsets for 2 speaker and 3 speakers, respectively. 
For the DIHARD II dataset, we conduct model adaptation on the Dev part and evaluate on the Test part.
The CALLHOME dataset comprises 8kHZ telephone-channel recordings, and the DIHARD II dataset contains 16kHZ recordings from a diverse range of sources.
To simplify the training setup, we upsampled the CALLHOME dataset to 16kHZ in our experiments.

\vspace{-0.8em}
\begin{table}[!ht]
\caption{Real dataset configuration. \# Spk and \# Utt represent the number of speakers and utterances, respectively. Ovl.(\%) corresponds to the overlap ratio. The numbers in the Duration column represents the minimum duration/maximum duration/average duration of each utterance. }
\label{tab:real data}
\centering
\scriptsize
\begin{tabular}{p{2.2cm}<{\centering}p{0.6cm}<{\centering}p{0.6cm}<{\centering}p{0.6cm}<{\centering}p{0.5cm}<{\centering}p{1.8cm}<{\centering}}
\hline
Dataset & Split & \# Spk & \# Utt & Ovl.~(\%) & Duration~(hrs) \\
\hline
\multirow{2}{*}{CALLHOME-2spk~\cite{callhome}} & Part 1 & 2 & 155 & 14.0 &  0.86/2.21/1.23  \\
&Part 2 & 2 & 148 & 13.1 & 0.88/2.23/1.20 \\
\hline
\multirow{2}{*}{CALLHOME-3spk~\cite{callhome}} & Part 1 & 3 & 61 & 19.6 & 0.95/6.35/2.07  \\
&Part 2 & 3 & 74 & 17.0 & 0.77/8.21/2.42 \\
\hline
\multirow{2}{*}{DIHARD II~\cite{ryant2019second}} & Dev & 1-10 & 192 & 9.8 & 0.45/11.62/7.44 \\
& Test & 1-9 & 194 & 8.9 & 0.63/13.50/6.96 \\
\hline
\end{tabular}
\end{table}

\vspace{-1em}
\subsection{Text Prompt Generation}
In the first scenario ``TSD with Semantic Description'' referred to Section~\ref{subsubsection:textual tsd}, we take three semantic attributes as examples in our work, denoted as \textbf{Gender}, \textbf{Speaker counter} and \textbf{Keynote speaker}.
The gender attribute contains two values, female and male, which can guide the system to output the gender-specific event regions.
Speaker counter attribute identifies the number of concurrent speakers at each frame and contains three event values: non-speech, single-speaker speech, and overlapped speech. 
Keynote speaker attribute focuses on identifying the keynote speaker. It contains one event value to represent the person who talks most. 

In practice, human's descriptions of an target event speech are often
diverse. To mimic the real-world scenario, for each target event, we first prepare a single command template. Then each template will undergo rephrasing and expansion through ChatGPT-4, resulting in the generation of 50 distinct text commands.
To be precise, we utilize an 80\%/10\%/10\% partitioning method that ensures non-overlapped training/validation/test sets. The text commands associated with each target event in the testing set remain unseen for those of the training process to ensure the generalization ability of our text encoder and our framework.

We show the ChatGPT command for ``male speech event'' as an example.
\begin{center}
\fcolorbox{black}{gray!10}{\parbox{.9\linewidth}{
\textbf{ChatGPT command: } \\
You are asked to come up with 50 diverse instructions rephrased and expanded from the template ``Please detect the regions that male speech occurs in the audio.'' \\ Here are the requirements: 
1. These instructions should be to instruct someone to identify the target event regions. \\
2. Try not to repeat the verb for each instruction to maximize diversity. \\
3. The type of instructions should be diverse. \\
4. The instructions should be oral English. \\
List of instructions:}}
\end{center}

Our generation method yields 50 instructions and some examples are shown.
\begin{center}
\fcolorbox{black}{gray!10}{\parbox{.9\linewidth}{
\textbf{Male Speech:}\\
1. I need you to identify areas in this recording where male voices are present.  \\
2. Your task is to find and label the instances of male dialogue in this recording. \\
3. Your objective is to identify the segments where men are speaking in this audio. \\
4. Can you mark out the sections where men's voices appear in this audio track? \\
5. Please trace the intervals in this sound clip featuring speech from a male.}}
\end{center}

\vspace{-1em}
\subsection{Evaluation Metrics}
The output of the target speech diarization system focuses on identifying each target event regions rather than all speakers' activities like traditional diarization systems. Therefore, as a new task, it's not appropriate to use the traditional speaker diarization metric. Thus,
we primarily employed three metrics: accurate precision~(AP), area under the receiver operating characteristic~(AUC), and equal error rate~(EER) based on the implementation from scikit-learn package.

Moreover, when the pre-enrolled speech of all speakers are provided, our system is functioned as the traditional speaker diarization. In this scenario, we report the diarization error rate~(DER) to show our effectiveness compared with other SOTA speaker diarization systems.

\vspace{-1em}
\section{Experimental Setup}
\subsection{Implementation Details}
\subsubsection{Training and Inference Details} The proposed MM-STD framework was implemented using PyTorch and optimized with the Adam optimizer. We set the initial learning rate to $10^{-4}$ and decrease it by 5\% for each epoch. 

To achieve a multi-task unified model, our study utilized the parallel characteristic of the Transformer decoder structure and adopted a multi-task training strategy. To fully explore all prompt-aligned input-label pairs, we provided all events' prompts for each utterance. This allowed our multi-task training model to accommodate a wide range of prompts during the evaluation phase. 
It is worth noting that to achieve unified yet independent multi-task learning, we applied an attention mask for Transformer decoder. This ensured that diverse prompt embeddings remain independent, with the exception of prompts requiring text and audio interaction, such as prompts under ``TSD with Audio-language Logical Controller'' scenario as introduced in Section~\ref{subsubsection:audio-text}. 

\subsubsection{Model Details} The Pre-trained Language Model (PLM) consists of 6 Transformer encoder blocks, each with 12 attention heads and 768 hidden dimensions. 
Both the dimension $D$ of audio feature $F_a$ and prompt embeddings $E$ were set to 192. 
The pre-trained audio prompt encoder ECAPA-TDNN is trained on VoxCeleb2 dataset. The pre-trained face encoder ResNet50 is pre-trained on the Glint360K dataset~\cite{an2021partial}.
The voice-face aligner module is composed of a 4-layer multi-layer perceptron (MLP). Each layer consists of a linear layer followed by a Gaussian error linear unit (GeLU). The output dimensions of each layer are 1024, 1024, 256, and 512, respectively.
For both Transformer encoder and decoder structure, 4-layer Transformer with 8 attention heads was applied.

\vspace{-1em}
\subsection{Data Augmentation}
During training, we perform speech and face augmentation for audio and visual prompts, to improve the diversity of training samples, thus the robustness of audio and visual prompt embedding.
\subsubsection{Speech Augmentation} We apply an online augmentation strategy with two datasets: the RIR dataset~\cite{ko2017study} and the MUSAN dataset~\cite{snyder2015musan}.
The RIR dataset contains room impulse responses that can be used to simulate the reverberation effects via convolution. These effects occur due to signal reflections bouncing off surfaces such as walls, floor, and other objects within an acoustic enclosure. Meanwhile, the MUSAN dataset~\cite{snyder2015musan} contains a variety of ambient sounds, including nature noises~(such as the sounds from train, thunder, rain), background music~(instrument or singing) and babble~(multi-speaker talking simultaneously).
\subsubsection{Face Augmentation} 
Facial images are usually distracted by non-identity information, such as colour, background, and image layout. A well-designed face augmentation approach can assist the encoder in capturing distinctive facial features more effectively.
Firstly, we align all the faces with the detected landmarks during pre-processing~\cite{zhang2016joint} since the unaligned faces in the training set make recognition harder~\cite{kowalski2017deep}.
We then reshape the face image into 3$\times$112$\times$112, and apply the random horizontal flip with probability 0.5.
Finally, we apply Gaussian blur techniques with kernel size 5$\sim$9 and sigma 0.1$\sim$5, and randomly convert image to gray scale with a probability of 0.2.

\vspace{-1em}
\section{Results and Analysis}
\begin{table*}[!t]
\caption{The results of MM-TSD trained on MM-2spk dataset and tested on both Seen-heard and Unseen-unheard test sets. The performance showcases the effectiveness of MM-TSD in detecting the target events guided by different types of prompts across diverse semantic attributes.}
\label{tab:main}
\centering
\begin{tabular}{c|c|c|c|ccc|ccc}
\hline
\multicolumn{3}{c|}{Prompt Modality} & \multirow{2}{*}{Attribute} & \multicolumn{3}{c|}{Seen-Heard} & \multicolumn{3}{c}{Unseen-Unheard}\\
\cline{1-3}
\cline{5-10}
Text & Audio & Visual & & AP~(\%)$\uparrow$&AUC~(\%)$\uparrow$&EER~(\%)$\downarrow$ & AP~(\%)$\uparrow$&AUC~(\%)$\uparrow$&EER~(\%)$\downarrow$\\
\hline
\multirow{3}{*}{\checkmark} &&& gender &99.26&99.38&2.48&99.83&99.84&1.33 \\
&&&counter&99.16&99.45&2.87&99.72&99.86&1.71 \\
&&&keynote&99.88&99.52&3.34&99.88&99.49&3.32 \\
\hline
& \checkmark && speaker id. &97.72&98.09&7.11&95.46&95.91&11.99 \\
\hline
&& \checkmark & face id. &92.46&93.99&13.51&85.74&88.28&20.82 \\
\hline
\multirow{2}{*}{\checkmark} & \multirow{2}{*}{\checkmark} && included id.&97.67&98.06&7.18&95.37&95.84&11.91 \\
&&&excluded id.&96.07&96.03&8.86&96.56&95.81&11.67 \\
\hline
\end{tabular}
\end{table*}

In this section, we begin with an overview evaluation and analysis of our proposed MM-TSD framework across various modalities and attributes, showcasing the effectiveness of our approach in target speech diarization.
Additionally, we conduct a comparative analysis and ablation studies for each modality prompts to further demonstrate the efficacy of each component of our approach. 
Moreover, it's worth noting that the applicability of our framework extends beyond target speech diarization. With attribute-aligned prompts, it can be utilized for tasks such as traditional speaker diarization, 
overlap speech detection, and gender diarization. Remarkably, the performance of our framework is comparable to specialist models dedicated to these individual tasks. These investigations highlight the robustness and versatility of our system.

\vspace{-0.8em}
\subsection{Overall Evaluation of MM-TSD}
During the training phase of our framework, we trained the MM-TSD model on the MM-2spk dataset using audio-visual-text prompts. The performance results, including AP, AUC, and EER, across diverse prompt modalities and attributes, are detailed in Table~\ref{tab:main}.

Indeed, the performance of our system in the text modality is particularly impressive. It excels on both the ``Seen-Heard'' and ``Unseen-Unheard'' datasets, achieving AP and AUC values that consistently surpass 99\%, with EER values remaining under 4\%. These outcomes serve as compelling evidence of our MM-TSD framework's capability to accurately identify desired event regions, guided by any provided text commands.

Also, the model's specialization in attributes related to speaker identity during training could lead to over-fitting on seen speakers, resulting in better performance on the ``Seen-Heard'' set. 
When the unified model primarily focuses on semantic attributes related to speaker identity, it might allocate more resources to learning speaker-specific patterns, potentially at the expense of other semantic concepts. This could explain the slightly lower performance on attributes not directly related to speaker identity on the ``Seen-Heard'' set. 

In the case of audio prompts and audio-text prompts, as depicted in the fourth and last two rows of Table~\ref{tab:main}, our system continues to shine by effectively detecting or excluding the target speaker based on the provided text as logical controller and pre-enrolled audio. The AP and AUC values exceed 96\%, and EER values are under 12\%, even for unseen speakers. This demonstrates the cross-modal interaction and generalization ability of our system, which remains robust in various scenarios.

In the challenging task of face-based target speaker detection, our system achieves AP and AUC values exceeding 90\% for ``Seen-Heard'' speakers. However, it faces greater difficulty with unseen speakers, resulting in slightly lower performance due to the limited information provided by a single face image. Nevertheless, these results confirm our system's capacity to discern the relationship between a target speaker's voice and face, albeit with a greater challenge when dealing with unseen individuals.

As a conclusion, our MM-TSD system can support diverse prompts to detect different type of desired speech events.

\vspace{-0.5em}
\subsection{Text Modality Analysis}
\subsubsection{Parameter-efficient Exploration}
For the text prompt, we employ parameter-efficient tuning methods for Pre-trained Language Model (PLM) as our text prompt encoder. The pre-trained text encoder DistilBERT can be fine-tuned and adapted to our tasks and language commands by adding only a few trainable parameters.
Specially, we study the effect of various parameter-efficient techniques, such as bottleneck~(bn) adapter~\cite{bapna2019simple} and LoRA~\cite{hu2022lora} on the MM-2spk dataset. 
Table~\ref{tab:lora} shows a performance comparison on the MM-2spk dataset between a frozen DistilBERT~(denoted as ``DistilBERT''), a frozen DistilBERT with bottleneck adapter tuning~(denoted as ``+ bn adapter'') and a frozen DistilBERT with LoRA tuning~(denoted as ``+ LoRA'').
The results indicate that a frozen DistilBERT with parameter-effecient tuning outperforms the original fixed DistilBERT model, and LoRA shows slightly better performance than bottleneck adapter. Therefore, we adopt LoRA as our chosen tuning technique for DistilBERT as our text prompt encoder.

\begin{table}[!ht]
\caption{Prameter-efficient exploration for three different text prompt encoders of MM-TSD trained on MM-2spk. `DistilBERT' means the text encoder is the frozen DistilBERT only, ``+ bn adapter'' and ``+ LoRA'' represent the frozen DistilBERT with additional bottleneck adapter and LoRA tuning, respectively.}
\label{tab:lora}
\centering
\begin{tabular}{m{1.6cm}<{\centering}|m{1.0cm}<{\centering}|m{0.9cm}<{\centering}m{0.9cm}<{\centering}m{0.9cm}<{\centering}m{0.9cm}<{\centering}}
\hline
Text Encoder & Attribute & ACC~(\%)$\uparrow$ & AP~(\%)$\uparrow$ & AUC~(\%)$\uparrow$ & EER~(\%)$\downarrow$ \\
\hline
\multirow{3}{*}{DistilBERT} & gender &97.78&99.53&99.56&2.26 \\
&counter& 94.48&97.19&97.93&6.36 \\
&keynote &98.28&99.95&99.81&2.10 \\
\hline
\multirow{3}{*}{+ bn adapter} & gender &98.09&99.67&99.71&1.96 \\
&counter &98.54&99.61&99.74&1.75 \\
&keynote &98.37&99.96&99.85&1.89 \\
\hline
\multirow{3}{*}{+ LoRA} & gender &\textbf{98.23}&99.63&99.70& 1.81\\
&counter &\textbf{98.76}&99.78&99.87&1.36 \\
&keynote &\textbf{98.40}&99.96&99.83&1.96 \\
\hline
\end{tabular}
\end{table}

\subsubsection{Variable Number of Speakers}
We further evaluate the performance of our system on MM-3spk and MM-4spk datasets, each featuring 3 and 4 speakers, respectively. The results are summarized in Table~\ref{tab:sim3spk}. For these challenging conditions with more speakers, both the AP and AUC of MM-TSD still surpass 99 \%, which demonstrate its ability to accurately detect the prompt-specified target event in conversation scenarios involving multiple speakers.

\begin{table}[!ht]
\caption{The results of MM-TSD trained on MM-3spk and MM-4spk datasets with text prompts.}
\label{tab:sim3spk}
\centering
\begin{tabular}{c|c|ccc}
\hline
Dataset & Attribute &AP~(\%)$\uparrow$& AUC~(\%)$\uparrow$&EER~(\%)$\downarrow$ \\
\hline
\multirow{3}{*}{MM-3spk} & gender &99.20&99.32&2.42 \\
&counter&99.75&99.87&1.56 \\
&keynote&99.88&99.60&3.18 \\
\hline
\multirow{3}{*}{MM-4spk} & gender &99.27&99.37&2.54 \\
&counter&99.53&99.69&2.17 \\
&keynote&99.88&99.62&3.07 \\
\hline
\end{tabular}
\end{table}

\subsubsection{MM-TSD v.s. OSD}
\begin{figure}[!ht]
    \centering
      \includegraphics[scale=0.45]{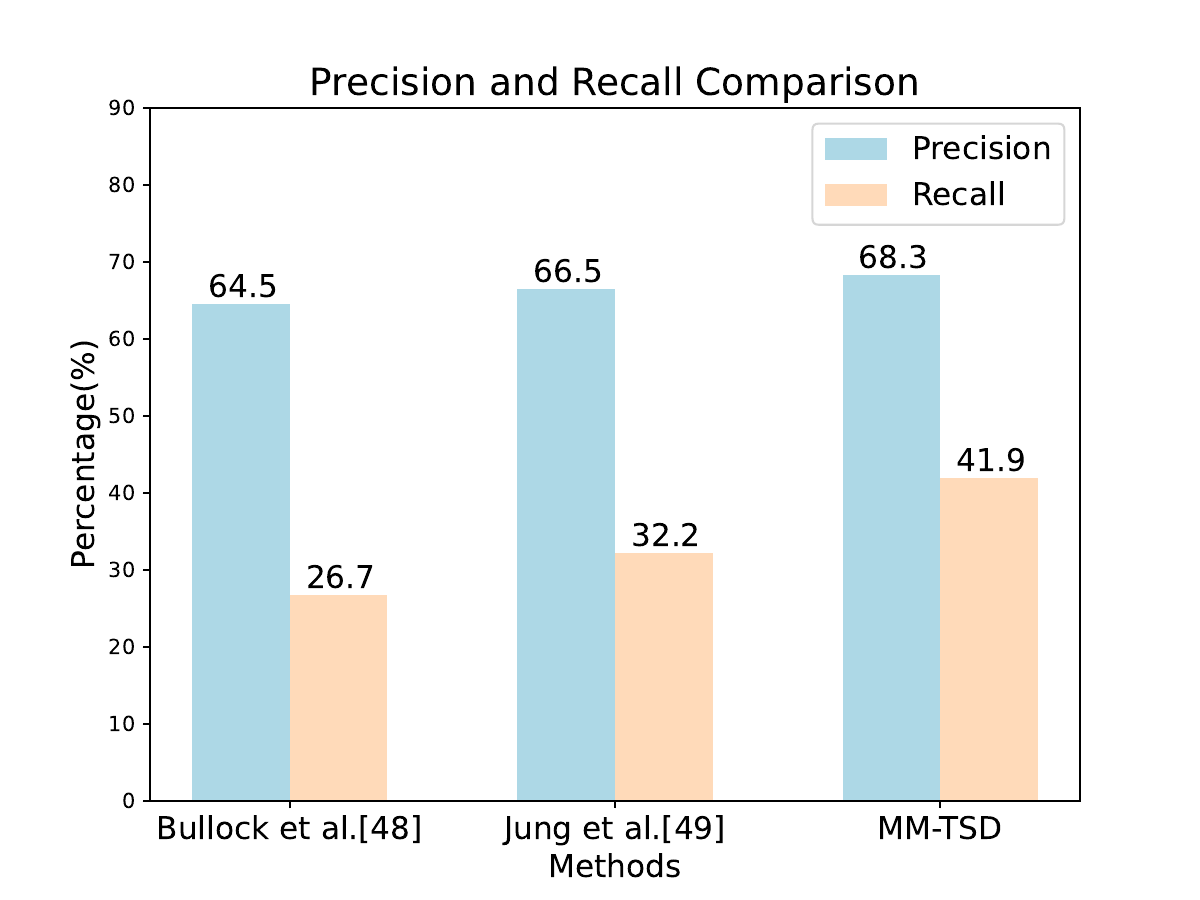}
    \vspace{-2em}
    \caption{The precision and recall results of MM-TSD when it achieves the overlap speech detection task on the DIHARD II evaluation set, and results of other specialized methods are also shown for comparison.}
    \label{fig:count}
    \vspace{-1em}
\end{figure}
MM-TSD can also be functioned as a three-class speaker counter, capable of estimating the number of concurrent speakers at each frame when we provide prompts for non-speech, single speaker speech and overlapped speech simultaneously.

We evaluate the performance of MM-TSD in the overlapped speech detection~(OSD) task~\cite{boakye2008overlapped} on DIHARD II evaluation set. We provide the text prompts aligned with the overlapped speech event, which was initially trained on the MM-4spk training set and further fine-tuned on the DIHARD II development set. Figure~\ref{fig:count} compares the results of MM-TSD with two previous OSD models, which are proposed by Bullock et al~\cite{bullock2020overlap} and Jung et al~\cite{jung2021three} respectively. MM-TSD achieves significantly better precision at $68.3\%$ and recall at $41.9\%$ compared to the specialized overlapped speech detection models on DIHARD II evaluation set.

\subsubsection{Structure Ablation}
We then investigated the Transformer encoder-decoder architecture within MM-TSD using text prompts. 
We implemented two baseline approaches that remove the Transformer decoder part and treated each semantic event detection as a supervised frame-wise classification problem.
The first baseline still use the Transformer encoder to obtain speech representation named as ``Transformer Encoder''. For the second baseline, we replaced the Transformer encoder with ECAPA-TDNN encoder, named as ``TDNN Encoder''. The encoder was followed by a linear layer for each event. 
In our MM-TSD approach, prompts were used to specify the target events, and a Transformer decoder decoded the event occurrence probability from Transformer encoder representations as key and value, and the prompt embedding as query, as illustrated in Figure~\ref{fig:transformer}.

The results reported in Table~\ref{tab:structure} were trained on the MM-4spk training set and validated on MM-4spk test set.
The good performance of ``Transformer Encoder'' as a baseline demonstrates that Transformer encoder representations contain sufficient information about gender, concurrent number of speakers, and speaker duration. 
Our MM-TSD achieves comparable performance with Transformer encoder-only baseline. 
MM-TSD utilizes prompt embeddings as queries inputted into the Transformer decoder to accurately retrieve the target event related information from Transformer encoder representations. Additionally, our framework can produce more flexible prompt-driven outputs with the Transformer decoder structure. 

\vspace{-0.5em}
\begin{table}[!ht]
\setlength{\tabcolsep}{7pt}
\caption{Structure ablation comparison between encode-only methods and prompt-driven encoder-decoder structure. The results are evaluated on MM-4spk test set. }
\label{tab:structure}
\begin{tabular}{m{1.7cm}
<{\centering}|m{0.8cm}
<{\centering}|m{0.9cm}<{\centering}m{0.9cm}<{\centering}m{0.9cm}<{\centering}m{0.9cm}<{\centering}}
\hline
Method & Attribute & 
ACC~(\%)$\uparrow$ &AP~(\%)$\uparrow$ & AUC~(\%)$\uparrow$ & EER~(\%)$\downarrow$\\ \hline
\multirow{3}{*}{TDNN Encoder} & gender & 97.09&98.95 & 99.17 & 2.94    \\ 
&counter &97.97&99.54&99.77&2.13 \\
& keynote &94.50&99.46&98.41&6.72 \\ \hline
\multirow{3}{*}{\begin{tabular}[c]{@{}l@{}}Transformer\\ Encoder\end{tabular}} & gender & 97.40 & 99.19 &99.32 & 2.69 \\ 
&counter &98.31&99.72&99.86&1.78 \\
& keynote &97.03&99.79&99.36&3.39 \\ \hline
\multirow{3}{*}{\textbf{MM-TSD}} &  gender & {97.51} &99.27 & 99.37 & 2.54    \\ 
&counter &98.02&99.53&99.69&2.17 \\
& keynote &{97.28}&99.88&99.62&3.07 \\ \hline
\end{tabular}
\vspace{-0.8em}
\end{table}

\vspace{-1.em}
\subsection{Audio Modality Analysis}
In the context of audio prompts, when pre-enrolled speeches from all speakers are available, our framework can function as a traditional speaker diarization system. To demonstrate the effectiveness of MM-TSD with audio prompts, we conducted a comparative analysis with traditional speaker diarization systems.

The main difference lies in the utilization of pre-enrolled speech as auxiliary information in MM-TSD with audio prompts from all speakers.
Inspired by TS-VAD~\cite{medennikov2020target}, we employed a clustering-based system~\cite{wang2023wespeaker} as a frontend to obtain reference speech for each speaker roughly, achieving 15.60\% and 21.25\% DER performance on Audio-2spk and Audio-3spk datasets, respectively. Subsequently, our MM-TSD system with estimated reference speech as audio prompt was applied to detect the activity of each speaker with the frontend pre-enrollment.

In Table~\ref{tab:sd}, we provide the DER performance on the CALLHOME Part 2 2-spk and 3-spk subsets. When oracle pre-enrolled speeches are provided for each speaker, our MM-TSD achieves impressive DER values of 6.80\% and 8.51\% for 2 and 3 speakers, respectively. Even after adopting a clustering-based system to obtain the estimated pre-enrolled speech for each speaker as audio prompts, our system obtains DER values of 8.17\% and 11.76\% for 2 and 3 speakers, respectively, and maintains comparable performance with various state-of-the-art speaker diarization systems. Moreover, our MM-TSD is a more general framework, and the others are all specialized systems.


\begin{table}[!ht]
\caption{DER~(\%) results comparison on the CALLHOME Part 2 2-spk and 3-spk subsets. Lower is better. `MM-TSD~($\mathcal{A}$)-oracle' represents MM-TSD with the ground-truth enrolled speech as audio prompts for speaker diarization, `MM-TSD~($\mathcal{A}$)-clust.' denotes that the audio prompts come from the estimated target speech by clustering front-end.}
\label{tab:sd}
\centering
\begin{threeparttable}
\begin{tabular}{p{2.7cm}<{\centering}|p{2.4cm}<{\centering}p{2.4cm}<{\centering}}
\hline
Method & CALLHOME~(2spk) & CALLHOME~(3spk)  \\
\hline
x-vector clustering~\cite{horiguchi2020end} & 11.53 & 19.01 \\
clutering frontend~\cite{wang2023wespeaker
} &15.60& 21.25 \\
BLSTM-EEND~\cite{fujita2019end} & 26.03 & -\\
SA-EEND~\cite{horiguchi2020end,fujita2019end} & 9.54 & 14.00 \\
SC-EEND~\cite{fujita2020neural} & 8.86 & - \\
EEND-EDA~\cite{horiguchi2020end} & 8.07 & 13.92 \\
EEND-EDA $^\dagger$ & 8.32 & 17.07 \\
TS-VAD~\cite{medennikov2020target} & 9.51 & -\\
MTFAD~\cite{medennikov2020target} & 7.82 & - \\
AED-EEND~\cite{czy-taslp} & 7.75 & 12.87 \\
\hline
MM-TSD~(oracle) & \textbf{6.80} & \textbf{8.51} \\
MM-TSD~(clustering) & \textbf{8.17} & \textbf{11.76} \\
\hline
\end{tabular}
\begin{tablenotes}\footnotesize
\item $^\dagger$: our implementation.
\end{tablenotes}
\end{threeparttable}
\vspace{-1.5em}
\end{table}

\vspace{-1.em}
\subsection{Visual Modality Analysis}
\subsubsection{Voice-face Alignment}
MM-TSD applies the face prompt for face-based target speech diarization. To study the robustness of our proposed voice-face alignment module, we presented comparative results with and without voice-face alignment for different gender pairs, showcasing the significant impact on performance for both the ``Seen-Heard'' and ``Unseen-Unheard'' speakers, as detailed in Table~\ref{tab:av-align}.

In the ``Seen-Heard'' test set, the target speakers are already present during training, making it relatively straightforward to match the speaker's facial features with its speaking regions. As a result, it is natural that the performance is consistently better than that on the `Unseen-Unheard' set. Furthermore, we observe that the results exhibit slight improvements after the incorporation of audio-visual alignment.

The ``Unseen-Unheard'' test sets consist of samples that were not encountered during the training of both the voice-face alignment and face-based diarization tasks. Although the discrimination performance of unseen people is slightly lower than that of seen pairs, it's worth noting that the performance remains significantly superior to the ``w/o align'' baseline.

Interestingly, results for different gender pairs, namely F-M and M-F sets, consistently demonstrate good performance. This is attributed to the distinct gender characteristics, which make it easier to select the target speech using the provided images.
Conversely, mixtures with the same gender, i.e., M-M and F-F sets, pose a greater challenge in terms of differentiation based solely on the given images. However, the performance improvements following voice-face alignment are notably more pronounced in these cases.
The results demonstrate that our voice-face alignment module indeed learn the intrinsic relationship of cross-modal biometrics, that will help face-based diarization performance. To the best of our knowledge, our paper is the first work to propose the face-based diarization and it can be used for the diverse scenarios such as meeting discussion and human-robot interaction.

\vspace{-1em}
\begin{table}[!ht]
\caption{Comparative study between the voice-face aligner and previous systems for cross-modal speaker verification. Results are performed on the different datasets.}
\label{tab:cross verif}
\centering
\begin{tabular}{c|ccc}
\hline
Method & EER~(\%) $\downarrow$ & AUC~(\%) $\uparrow$ & Dataset \\
\hline
DIMNet~\cite{wen2018disjoint} & 24.56 & NA & VoxCeleb, VGGFace \\
SSNet~\cite{nawaz2019deep} & 29.50 & 78.8 & VoxCeleb \\
Pins~\cite{nagrani2018learnable} & 29.60 & 78.5 & VoxCeleb \\
\textbf{VF Aligner~(Ours)}&\textbf{24.40}&\textbf{83.0}& VoxCeleb2 \\
\hline
\end{tabular}
\end{table}

\subsubsection{Cross-modal Verification}
Face-based diarization relies on the ability of cross-modal speaker verification, which determines if a given voice segment and face image belong to the same person~\cite{nawaz2021cross,tao2020audio}. So we further evaluated the performance of our Voice-Face Aligner~(VF Aligner) on VoxCeleb2 dataset for this task and compared it with some of the existing systems. The results are presented in Table~\ref{tab:cross verif}, with Equal Error Rate~(EER) and Area Under the Curve~(AUC) as performance metrics. 

We observe that our audio-visual module with voice-face aligner effectively performs for cross-modal speaker verification.  
After alignment, our audio-visual module can perform comparably to the existing systems for cross-modal verification. This highlights the effectiveness of our voice-face aligner module in capturing the general higher-order information between voice characteristic and face appearance, including but not limited to age, nationality and gender.

\begin{table*}[t]
\caption{The results of MM-TSD for face-based diarization, different gender combinations with or without voice-face alignment are studied. Each test sample contains two speakers, `M' denotes male and `F' denotes the female.}
\label{tab:av-align}
\centering
\begin{tabular}{c|c|ccc|ccc}
\hline
\multirow{2}{*}{Visual Encoder} & \multirow{2}{*}{Gender} & \multicolumn{3}{c|}{Seen-heard Speaker} & \multicolumn{3}{c}{Unseen-unheard Speaker} \\
\cline{3-8}
&& AP~(\%)$\uparrow$ & AUC~(\%)$\uparrow$ & EER~(\%)$\downarrow$ & AP~(\%)$\uparrow$ & AUC~(\%)$\uparrow$ & EER~(\%)$\downarrow$ \\
\hline
\multirow{4}{*}{w/o align} & M-M &97.63&97.84&7.73& 77.11&79.73&29.09 \\
& M-F &99.80&99.82&1.98&99.90&99.91&1.52 \\
& F-M &99.84&99.84&1.74&96.37&97.15&7.03 \\
& F-F &94.75&95.63&10.78&57.54&62.59&40.20 \\
\hline
\multirow{4}{*}{with align} & M-M &{98.05}&{98.27}&{6.59}
&{84.00}&{85.81}&{22.59} \\
& M-F &99.84&99.85&1.85&99.90&99.91&1.39 \\
& F-M &99.89&99.89&1.28&96.71&96.95&6.72 \\
& F-F &{96.60}&{97.04}&{9.49}&{74.52}&{77.69}&{27.68} \\
\hline
\end{tabular}
\vspace{-1em}
\end{table*}

\vspace{-0.5em}
\subsection{Visualization of Results on Real-World Data}
\begin{figure*}[!t]
    \centering
      \includegraphics[scale=0.65]{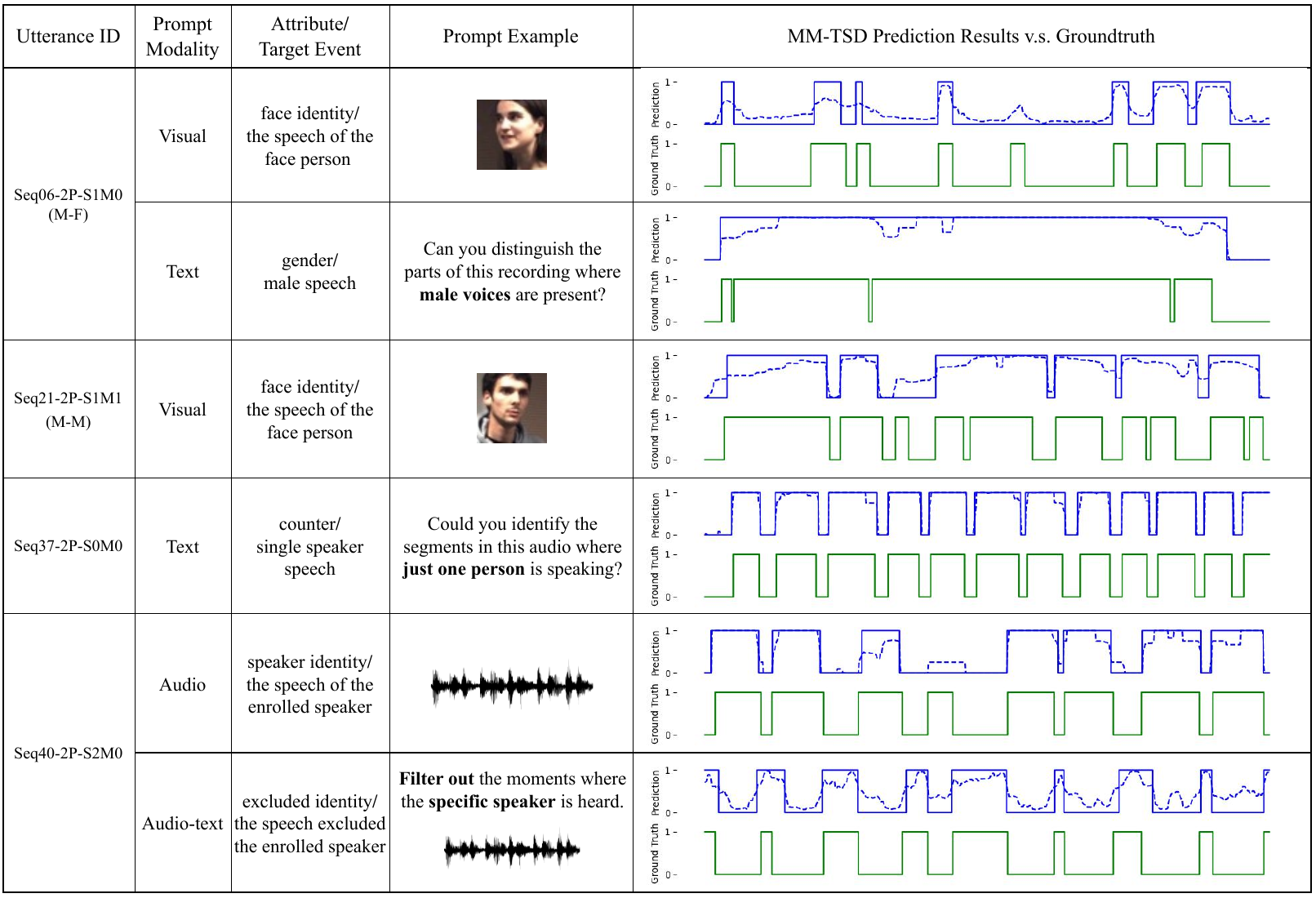}
    \caption{Visualization of results on real world samples selected from AVDIAR dataset. Each row represents an utterance id in AVDIAR dataset, followed by the prompt modality, semantic attribute, target event, prompt example, and visualization comparison of prediction results versus groundtruth.
    The green solid lines depict groundtruth labels of the target event, while the blue dashed lines represent the probabilities of the target event's occurrence at each frame. The blue solid lines indicate the model's prediction output of the target event.}
    \label{fig:vis}
\end{figure*}

In previous experimental sessions, we showcased the effectiveness of each modality module within MM-TSD and evaluated them separately on real datasets. 
To further highlight the unified capability of all modalities to handle complex conversations in real-world data, we conducted an assessment on the AVDIAR (Audio-Visual Diarization) dataset~\cite{gebru2018audiovisual}. This dataset is dedicated to the audio-visual analysis of real-world conversational scenes. Due to the absence of labels for all potential target events that MM-TSD can support, we manually labeled some samples and tested MM-TSD trained on MM-2spk dataset without further fine-tuning. Additionally, for an intuitive understanding of our framework's functionality, we provide visualizations of some prompt examples and their corresponding prediction results from our proposed MM-TSD framework in Figure~\ref{fig:vis}. 

As our prompt examples show, when we provide a text command like ``Can you distinguish the parts of this recording where male voices are present?", our framework will output the male speech regions within the input utterance audio~(the second row). In the scenario where the desired speaker's face is available, our framework can detect the active speaking regions according to this face identity. It's reasonable to expect that utterances with different genders~(the first row) will perform better than those with the same gender~(the third row), which is consistent with our previous analysis results. Moreover, when the target speaker's pre-enrolled speech is available, our framework can detect all the speaking regions accordingly. Furthermore, if we want to ``Filter out the moments where the specific speaker is heard'', we can provide the target speech and text commands so that our framework can achieve a NOT Gate function and output the moments without the enrolled speaker's voice.

\vspace{-1em}
\subsection{Future Work and Discussion}
\textbf{Complementary and composite prompts.} Currently, our framework supports textual, audio, or visual prompts to determine target events. In the scenario involving audio-text prompts, the text merely functions as the logical controller. However, the potential of audio-visual-textual prompts as complementary inputs could be explored in future research. For instance, to specify female speech, we could provide a text prompt for female speech regions, along with the female face image and enrollment female speech simultaneously. To achieving this, multi-modal training requires aligned audio-visual-textual data collection.

Furthermore, composite prompt-specified events, such as ``the female single speaker speech," could be explored in the future. Currently, in our framework, we would need to separate such composite events into single events ``female speech" and ``single speaker speech". Then, using two prompts to output two event regions, we can compute the composite event region. We aim to extend our framework to handle such composite prompts directly in our future work.

\textbf{Broader scope.} While our study focuses on three semantic attributes as examples for natural language descriptions of target speech, the potential scope of semantic description is much broader. Attributes such as age, pitch, nationality, and others could provide valuable context for specifying target speech events. Although our paper introduces a foundational model, it inherently cannot cover the entire spectrum of attribute scope. However, given access to relevant data, our proposed framework and training strategy can be adapted to accommodate any semantic attributes, offering a more comprehensive and adaptable solution.

Moreover, our current framework operates in a supervised manner, relying on labeled data for training. Expanding its capability to handle out-of-domain events, where labeled data may be scarce or unavailable, would significantly enhance its utility and effectiveness. Developing techniques for unsupervised or semi-supervised learning within our framework could unlock its potential to address a wider range of real-world scenarios and applications.

\vspace{-1em}
\section{Conclusion}
Our paper proposed the new task named target speech diarization to detect ``when target event occurs''. To solve this problem, we have introduced MM-TSD, a foundational model for target speech diarization that supports diverse and multi-modal prompts. MM-TSD enables users to utilize semantic language descriptions, pre-enrollment speech, pre-registered face images and audio-language logical prompts to specify the target event(s). We have conducted a comprehensive evaluation, including an overview and analysis of each modality, across various tasks such as traditional speaker diarization and overlap speech detection. MM-TSD achieved comparable performance with state-of-the-art specialist models. Furthermore, we made the first attempt at face-based target speech diarization with voice-face alignment. Our results demonstrate that MM-TSD is a promising approach for effectively addressing the target speech diarization problem. Our framework can be applied for human-robot interaction and meeting analysing scenarios. In future work, we plan to explore complementary prompts and extend the scope of our framework to further enhance its comprehensiveness.

\bibliographystyle{IEEEtran}
\bibliography{refs.bib}
\end{document}